\documentclass[english]{IEEEtran}
\usepackage[T1]{fontenc}
\usepackage[latin9]{inputenc}
\usepackage[active]{srcltx}
\usepackage{amsmath}
\usepackage{amssymb}
\usepackage{graphicx}

\makeatletter

\providecommand{\tabularnewline}{\\}

\author{
    \IEEEauthorblockN{Abubakr O. Al-Abbasi\IEEEauthorrefmark{1}, Ridha Hamila\IEEEauthorrefmark{1}, Waheed U. Bajwa\IEEEauthorrefmark{2}, and Naofal Al-Dhahir\IEEEauthorrefmark{3}}

    \IEEEauthorblockA{\IEEEauthorrefmark{1}
Dept. of Electrical Engineering, Qatar University, Qatar
}

\IEEEauthorblockA{\IEEEauthorrefmark{2}
Dept. of Electrical and Computer Engineering, Rutgers University, USA
}

    \IEEEauthorblockA{\IEEEauthorrefmark{3}
Dept. of Electrical Engineering, University of Texas at Dallas, USA
}

\thanks{This paper was made possible by grant number NPRP 06-070-2-024 from
the Qatar National Research Fund (a member of Qatar Foundation). The
statements made herein are solely the responsibility of the authors.}
}

\usepackage{balance} 

\makeatother

\usepackage{babel}
\begin{document}

\title{A General Framework for the Design and Analysis of Sparse FIR Linear
Equalizers}
\maketitle
\begin{abstract}
Complexity of linear finite-impulse-response (FIR) equalizers is proportional
to the square of the number of nonzero taps in the filter. This makes
equalization of channels with long impulse responses using either
zero-forcing or minimum mean square error (MMSE) filters computationally
expensive. Sparse equalization is a widely-used technique to solve
this problem. In this paper, a general framework is provided that
transforms the problem of sparse linear equalizers (LEs) design into
the problem of sparsest-approximation of a vector in different dictionaries.
In addition, some possible choices of sparsifying dictionaries in
this framework are discussed. Furthermore, the worst-case coherence
of some of these dictionaries, which determines their sparsifying
strength, are analytically and/or numerically evaluated. Finally,
the usefulness of the proposed framework for the design of sparse
FIR LEs is validated through numerical experiments.
\end{abstract}

\vspace{-1em}

\section{Introduction \label{sec:Introduction}}

In numerous signal processing applications such as equalization and
interference cancellation, long FIR filters have to be implemented
at high sampling rates. This results in high complexity, which grows
proportional to the square of the number of nonzero taps. One approach
to reduce this complexity is to implement only the most significant
FIR filter taps, i.e., sparse filters. However, reliably determining
the locations of these dominant taps is often very challenging. 

Several design approaches have been investigated in the literature
to reduce the complexity of long FIR filters. In \cite{strongestTap},
the number of nonzero coefficients is reduced by selecting only the
significant taps of the equalizer. Nonetheless, knowledge of the whole
equalizer tap vector is required which increases the computational
complexity. In \cite{linearProgAlnOppen10}, an $\ell_{1}$-norm minimization
problem is formulated to design a sparse filter. However, since the
resulting filter taps are not exactly sparse, a strict thresholding
step is required to force some of the nonzero taps to $0$. An algorithm,
called sparse chip equalizer, for finding the locations of sparse
equalizer taps is given in \cite{sparseChipEqu} but this approach
assumes that the channel itself is sparse. In \cite{sparseFilterDesign13},
a general optimization problem for designing a sparse filter is formulated
that involves a quadratic constraint on filter performance. Nonetheless,
the number of iterations of the proposed backward selection algorithm
becomes large as the desired sparsity of the filter increases. In
addition, the approach in \cite{sparseFilterDesign13} also involves
inversion of a large matrix in the case of long Channel Impulse Responses
(CIRs). In \cite{newDFW}, a framework for designing sparse FIR equalizers
is proposed. Using greedy algorithms, the proposed framework achieved
better performance than just choosing the largest taps of the MMSE
equalizer, as in \cite{strongestTap}. However, this approach involves
Cholesky factorization, whose computational cost could be large in
the case of channels with large delay spreads. In addition, no theoretical
guarantees are provided.

In this paper, we develop a general framework for the design of sparse
FIR equalizers that transforms the original problem into one of sparse
approximation of a vector using different dictionaries. The developed
framework can then be used to find the sparsifying dictionary that
leads to the sparsest FIR filter subject to an approximation constraint.
We also investigate the coherence of the sparsifying dictionaries
that we propose as part of our analysis and identify one that has
the smallest coherence. Then, we use simulations to validate that
the dictionary with the smallest coherence gives the sparsest FIR
linear equalizer. Moreover, the numerical results demonstrate the
significance of our approach compared to conventional sparse FIR equalizers
(e.g., \cite{strongestTap}) in terms of both performance and computational
complexity.

\textbf{\textit{\small{}Notations}}: We use the following standard
notation in this paper: $\mbox{\ensuremath{\boldsymbol{I}}}_{N}$
denotes the identity matrix of size $N$. Upper and lower case bold
letters denote matrices and vectors, respectively. The notations $(.)^{-1},\,(.)^{*}\mbox{ and }\,(.)^{H}$
denote the matrix inverse, the matrix (or element) complex conjugate
and the complex-conjugate transpose operations, respectively. $\mbox{E\ensuremath{\left[.\right]}}$
denotes the expected value operator. The components of a vector starting
from $k_{1}$ and ending at $k_{2}$ are given as subscripts to the
vector separated by a colon, i.e., $\boldsymbol{x}_{k_{1}:k_{2}}.$ 

\vspace{-1.0em}

\section{System Model\label{sub:Signal-Model}}

A linear, time invariant, dispersive and noisy communication channel
is considered. The standard complex-valued equivalent baseband signal
model is assumed. At time $k$, the received sample $y_{k}$ can be
expressed as 

\vspace{-1.0em}

\begin{equation}
y_{k}=\sum_{l=0}^{v}h_{l}\,x_{k-l}\,+n_{k},\label{eq:y_k}
\end{equation}
where $h_{l}$ is the CIR whose memory is $v$, $n_{k}$ is the additive
noise symbol and $x_{k-l}$ is the transmitted symbol at time ($k-l$).
At any time $k$, an FIR filter of length $N_{f}$ is applied to the
received samples in order to recover the transmitted symbols with
some possible time delay. For simplicity, we assume a symbol-spaced
equalizer but our proposed design framework can be easily extended
to the general fractionally-spaced case. For these $N_{f}$-long received
samples of interest, the input-output relation in (\ref{eq:y_k})
can be written compactly as 

\vspace{-1.2em}

\begin{equation}
\boldsymbol{y}_{k:k-N_{f}+1}=\boldsymbol{H}\,\boldsymbol{x}_{k:k-N_{f}-v+1}+\boldsymbol{n}_{k:k-N_{f}+1}\,,\label{eq:y_Hx_n}
\end{equation}
where $\boldsymbol{y}_{k:k-N_{f}+1},\,\boldsymbol{x}_{k:k-N_{f}-v+1}$
and $\boldsymbol{n}_{k:k-N_{f}+1}$ are column vectors grouping the
received, transmitted and noise samples. Additionally, $\boldsymbol{H}$
is an $N_{f}\times(N_{f} + \nu)$ Toeplitz matrix whose first row
is formed by $\{h_{l}\}_{l=0}^{l=v}$ followed by zero entries. It
is useful, as will be shown in the sequel, to define the output auto-correlation
and the input-output cross-correlation matrices based on the block
of length $N_{f}$. Using (\ref{eq:y_Hx_n}), the input correlation
and the noise correlation matrices are, respectively, defined by {\small{}$\boldsymbol{R}_{xx}\triangleq E\left[\boldsymbol{x}_{k:k-N_{f}-v+1}\boldsymbol{x}_{k:k-N_{f}-v+1}^{H}\right]\mbox{ and }\boldsymbol{R}_{nn}\triangleq E\left[\boldsymbol{n}_{k:k-N_{f}+1}\boldsymbol{n}_{k:k-N_{f}-1}^{H}\right]$}.
Both the input and noise processes are assumed to be white; hence,
their auto-correlation matrices are assumed to be (multiples of) the
identity matrix, i.e., $\boldsymbol{R}_{xx}=\boldsymbol{I}_{N_{f}+v}$
and $\boldsymbol{R}_{nn}=\frac{1}{SNR}\boldsymbol{I}_{N_{f}}$. Moreover,
the output-input cross-correlation and the output auto-correlation
matrices are, respectively, defined as

\vspace{-1.0em}

{\small{}
\begin{eqnarray}
\boldsymbol{R}_{yx} & \!\!\triangleq & \!\!E\left[\boldsymbol{y}_{k:k-N_{f}+1}\boldsymbol{x}_{k:k-N_{f}-v+1}^{H}\right]=\boldsymbol{H}\boldsymbol{R}_{xx}\,,\,\mbox{and}\\
\boldsymbol{R}_{yy} & \!\!\triangleq & \!\!E\left[\boldsymbol{y}_{k:k-N_{f}+1}\boldsymbol{y}_{k:k-N_{f}+1}^{H}\right]\!\!=\boldsymbol{H}\boldsymbol{R}_{xx}\boldsymbol{H}^{H}+\boldsymbol{R}_{nn}.\label{eq:R_yy_def}
\end{eqnarray}
}\vspace{-2em}

\section{Sparse FIR Linear Equalizers Design\label{sec:Sparse-FIR-Equalization}}

\subsection{Initial formulation}

The received samples are passed through an FIR filter with length
$N_{f}$. Hence, the error symbol at time $k$ is given by

\vspace{-1.2em}

\begin{equation}
e_{k}=x_{k-\Delta}-\hat{x}_{k}=x_{k-\Delta}-\boldsymbol{w}^{H}y_{k:k-N_{f}+1}\,,
\end{equation}
where $\Delta$ is the decision delay, typically {\small{}$0\leq\Delta\leq N_{f}+v-1$},
and $\boldsymbol{w}$ denotes the equalizer taps vector whose dimension
is $N_{f}\times1$. Using the orthogonality principle of linear least-squares
estimation, the MSE, denoted as $\xi\left(\boldsymbol{w}\right)$,
equals \cite{newDFW}

\vspace{-1em}

\begin{eqnarray*}
\xi\left(\boldsymbol{w}\right) & \ensuremath{\triangleq} & E\left[\left|e_{k}^{2}\right|\right]=\varepsilon_{x}-\boldsymbol{w}^{H}\boldsymbol{R}_{yx}-\boldsymbol{R}_{yx}^{H}\boldsymbol{w}+\boldsymbol{w}^{H}\boldsymbol{R}_{yy}\boldsymbol{w}\,,
\end{eqnarray*}
where $\varepsilon_{x}\triangleq E\left[x_{k-\Delta}^{2}\right]$.
By writing $x_{k-\Delta}=\boldsymbol{1}_{\triangle}^{H}x_{k:k-N_{f}-v+1}$
and $r_{\Delta}=\boldsymbol{R}_{yx}\boldsymbol{1}_{\Delta}$, where
$\boldsymbol{1}_{\Delta}$ denotes $\left(N_{f}+v\right)$-dimensional
vector that is zero everywhere except in the $(\Delta+1)$-th element
where it is one, it follows that

\vspace{-1.2em}

{\small{}
\begin{eqnarray}
\!\!\!\!\!\!\!\!\xi\left(\boldsymbol{w}\right)\!\!\!\!\!\! & = & \!\!\!\!\!\!\underbrace{\varepsilon_{x}-r_{\Delta}^{H}\boldsymbol{R}_{yy}^{-1}r_{\Delta}}_{\xi_{m}}+\underbrace{(\boldsymbol{w}-\boldsymbol{R}_{yy}^{-1}r_{\Delta})^{H}\boldsymbol{R}_{yy}(\boldsymbol{w}-\boldsymbol{R}_{yy}^{-1}r_{\Delta})}_{\xi_{e}(\boldsymbol{w})}.\label{eq:MSE}
\end{eqnarray}
}Since $\xi_{m}$ does not depend on $\boldsymbol{w}$, the MSE $\xi\left(\boldsymbol{w}\right)$
is minimized by minimizing the term $\xi_{e}(\boldsymbol{w})$. Hence,
the optimum selection for $\boldsymbol{w}$, in the MMSE sense, is
the well-known Wiener solution $\boldsymbol{w}_{opt}=\boldsymbol{R}_{yy}^{-1}\boldsymbol{r}_{\Delta}$.
However, in general, this optimum choice is undesirable since $\boldsymbol{w}_{opt}$
is not sparse and its implementation complexity increases proportional
to $(N_{f})^{2}$ which can be computationally expensive \cite{DCProakis}.
However, any choice for $\boldsymbol{w}$ other than $\boldsymbol{w}_{opt}$
increases $\xi_{e}(\boldsymbol{w})$, which leads to performance loss.
This suggests that we can use the excess error $\xi_{e}(\boldsymbol{w})$
as a design constraint to achieve a desirable performance-complexity
tradeoff. Specifically, we formulate the following problem for the
design of sparse FIR equalizers:

\vspace{-1.8em}

\begin{eqnarray}
\widehat{\boldsymbol{w}}_{s} & \triangleq & \underset{\boldsymbol{w}\in\mathbb{C}^{N_{f}}}{\mbox{arg}\mbox{min}}\,\,\left\Vert \boldsymbol{w}\right\Vert _{0}\,\,\,\,\mbox{subject to}\,\,\,\,\,\xi_{e}(\boldsymbol{w})\leq\delta_{eq}\,,\label{eq:opt_prob1}
\end{eqnarray}
where $\left\Vert \boldsymbol{w}\right\Vert _{0}$ is the number of
nonzero elements in its argument, $\left\Vert .\right\Vert _{2}$
denotes the $\ell_{2}$-norm and $\delta_{eq}$ can be chosen as a
function of the noise variance. While one can attempt to use convex-optimization-based
approaches (after replacing $\left\Vert .\right\Vert _{0}$ with its
convex approximation $\left\Vert .\right\Vert _{1}$ in (\ref{eq:opt_prob1})
to reduce the search space and to make it more tractable \cite{justRelax06})
in order to estimate the sparse approximation vector $\widehat{\boldsymbol{w}}_{s}$,
there exists a number of greedy algorithms with low complexity that
can be used in an efficient manner. Starting with this initial formulation,
we now discuss a general framework for sparse FIR LEs design such
that the performance loss does not exceed a predefined limit. 

\vspace{-1.2em}

\subsection{Proposed sparse approximation framework}

Unlike earlier works, including the one by one of the co-authors \cite{newDFW},
we provide a general framework for designing sparse FIR linear equalizers
that can be considered as the problem of sparse approximation using
different dictionaries. Mathematically, this framework poses the problem
of sparse FIR equalizers design as follows:

\vspace{-1.2em}

\begin{equation}
\!\widehat{\boldsymbol{w}}_{s}\triangleq\underset{\boldsymbol{w}\in\mathbb{C}^{N_{f}}}{\mbox{\mbox{arg}\mbox{min}}}\,\left\Vert \boldsymbol{w}\right\Vert _{0}\,\,\,\mbox{subject to}\,\,\,\left\Vert \boldsymbol{A}\left(\boldsymbol{\varPhi}\boldsymbol{w}-\boldsymbol{b}\right)\right\Vert _{2}^{2}\leq\delta_{eq}\,,\label{eq:propFW}
\end{equation}
where $\boldsymbol{\varPhi}$ is the dictionary that will be used
to sparsely approximate $\boldsymbol{b}$, while $\boldsymbol{A}$
is a known matrix and $\boldsymbol{b}$ is a known data vector, both
of which change depending upon the sparsifying dictionary $\boldsymbol{\varPhi}$.
Note that by completing the square in (\ref{eq:opt_prob1}), the problem
reduces to the one shown in (\ref{eq:propFW}). Hence, one can use
any decomposition for $\boldsymbol{R}_{yy}$ to come up with a sparse
approximation problem. By writing the Choleskey or eigen-value decompositions
for $\boldsymbol{R}_{yy}$, we can have different choices for $\boldsymbol{A}$,
$\boldsymbol{\varPhi}$ and $\boldsymbol{b}$. Some of these possible
choices are shown in Table \ref{tab:Examples-of-different}.
\begin{table}
\vspace{-2em}

{\footnotesize{}\protect\caption{{\footnotesize{}Examples of different sparsifying dictionaries.}\label{tab:Examples-of-different}}
}{\footnotesize \par}

\vspace{-1.0em}

{\footnotesize{}}%
\begin{tabular}[b]{|l|l|l|l|l|l|}
\hline 
\multicolumn{3}{|c|}{{\scriptsize{}Cholesky Factorization }} & \multicolumn{3}{c|}{{\scriptsize{}Eigen Decomposition}}\tabularnewline
\hline 
\multicolumn{3}{|c|}{{\scriptsize{}$\boldsymbol{R}_{yy}=\boldsymbol{L}\boldsymbol{L}^{H}$
or $\boldsymbol{R}_{yy}=\boldsymbol{P}\boldsymbol{\Lambda}\boldsymbol{P}^{H}$}} & \multicolumn{3}{c|}{{\scriptsize{}$\boldsymbol{R}_{yy}=\boldsymbol{U}\boldsymbol{D}\boldsymbol{U}^{H}$}}\tabularnewline
\hline 
{\scriptsize{}$\boldsymbol{A}$} & {\scriptsize{}$\boldsymbol{\varPhi}$} & {\scriptsize{}$\boldsymbol{b}$} & {\scriptsize{}$\boldsymbol{A}$} & {\scriptsize{}$\boldsymbol{\varPhi}$} & {\scriptsize{}$\boldsymbol{b}$}\tabularnewline
\hline 
\hline 
{\scriptsize{}$\boldsymbol{I}$} & {\scriptsize{}$\boldsymbol{L}^{H}$} & {\scriptsize{}$\boldsymbol{L}^{-1}\boldsymbol{r}_{\Delta}$} & {\scriptsize{}$\boldsymbol{I}$} & {\scriptsize{}$\boldsymbol{D}^{\frac{1}{2}}\boldsymbol{U}^{H}$} & {\scriptsize{}$\boldsymbol{D}^{-\frac{1}{2}}\boldsymbol{U}^{H}\boldsymbol{r}_{\Delta}$}\tabularnewline
\hline 
{\scriptsize{}$\boldsymbol{L}^{-1}$} & {\scriptsize{}$\boldsymbol{R}_{yy}$} & {\scriptsize{}$\boldsymbol{r}_{\Delta}$} & {\scriptsize{}$\boldsymbol{D}^{-\frac{1}{2}}\boldsymbol{U}^{H}$} & {\scriptsize{}$\boldsymbol{R}_{yy}$} & {\scriptsize{}$\boldsymbol{r}_{\Delta}$}\tabularnewline
\hline 
{\scriptsize{}$\boldsymbol{I}$} & {\scriptsize{}$\boldsymbol{\Lambda}^{\frac{1}{2}}\boldsymbol{\boldsymbol{P}}^{H}$} & {\scriptsize{}$\boldsymbol{\Lambda}^{-\frac{1}{2}}\boldsymbol{P}^{-1}\boldsymbol{r}_{\Delta}$} & {\scriptsize{}$\boldsymbol{D}^{\frac{1}{2}}$} & {\scriptsize{}$\boldsymbol{U}^{H}$} & {\scriptsize{}$\boldsymbol{D}^{-1}\boldsymbol{U}^{H}\boldsymbol{r}_{\Delta}$}\tabularnewline
\hline 
\end{tabular}{\footnotesize \par}

\vspace{-2.5em}
\end{table}
Note that the framework parameters (i.e., $\boldsymbol{A}$, $\boldsymbol{\varPhi}$
and $\boldsymbol{b}$ ) in the left list of Table \ref{tab:Examples-of-different}
result by defining the Cholesky factorization \cite{matAnalysis}
either in the form $\boldsymbol{R}_{yy}\triangleq\boldsymbol{L}\boldsymbol{L}^{H}$
or $\boldsymbol{R}_{yy}\triangleq\boldsymbol{P}\boldsymbol{\Lambda}\boldsymbol{P}^{H}$
(where $\boldsymbol{L}$ is a lower-triangular matrix, $\boldsymbol{P}$
is a lower-unit-triangular (unitriangular) matrix and $\boldsymbol{\Lambda}$
is a diagonal matrix). On the other hand, the columns on the right
result by letting $\boldsymbol{R}_{yy}\triangleq\boldsymbol{U}\boldsymbol{D}\boldsymbol{U}^{H}$,
where $\boldsymbol{U}$ is a unitary matrix whose columns are the
eigenvectors of the matrix $\boldsymbol{R}_{yy}$ and $\boldsymbol{D}$
is a diagonal matrix with the corresponding eigenvalues on the diagonal.
For instance, by assuming $\boldsymbol{L}^{H}$, $\boldsymbol{D}^{\frac{1}{2}}\boldsymbol{U}^{H}$
and $\boldsymbol{R}_{yy}$ as sparsifying dictionaries, the problem
in (\ref{eq:propFW}) can, respectively, take one of the forms shown
below

\vspace{-1.5em}

{\small{}
\begin{eqnarray}
 & \!\!\!\!\underset{\boldsymbol{w}\in\mathbb{C}^{N_{f}}}{\mbox{min}}\left\Vert \boldsymbol{w}\right\Vert _{0}\mbox{\,\,\,\,\mbox{s.t }\,\,\,\,\ensuremath{\left\Vert \left(\boldsymbol{L}^{H}\boldsymbol{w}-\boldsymbol{L}^{-1}\boldsymbol{r}_{\Delta}\right)\right\Vert _{2}^{2}\leq\delta_{eq}\,},}\\
 & \!\!\!\!\!\!\!\!\!\!\!\,\underset{\boldsymbol{w}\in\mathbb{C}^{N_{f}}}{\mbox{min}}\left\Vert \boldsymbol{w}\right\Vert _{0}\mbox{\,\,\mbox{s.t }\,\,\ensuremath{\left\Vert \left(\boldsymbol{D}^{\frac{1}{2}}\boldsymbol{U}^{H}\boldsymbol{w}-\boldsymbol{D}^{-\frac{1}{2}}\boldsymbol{U}^{H}\boldsymbol{r}_{\Delta}\right)\right\Vert _{2}^{2}\leq\delta_{eq},\,\mbox{and}}}\\
 & \!\!\!\!\!\!\underset{\boldsymbol{w}\in\mathbb{C}^{N_{f}}}{\mbox{min}}\left\Vert \boldsymbol{w}\right\Vert _{0}\mbox{\,\,\,\,\mbox{s.t }\,\,\,\,\ensuremath{\left\Vert \boldsymbol{L}^{-1}\left(\boldsymbol{R}_{yy}\boldsymbol{w}-\boldsymbol{r}_{\Delta}\right)\right\Vert _{2}^{2}}\ensuremath{\leq\delta_{eq}}\,}.
\end{eqnarray}
}{\small \par}

Note that we can reduce the decomposition complexity by approximating,
for reasonably large $N_{f}$, the Toeplitz $\boldsymbol{R}_{yy}$
by a circulant matrix whose eigenvectors are the Discrete Fourier
Transform (DFT) vectors and eigenvalues are the output discrete spectrum
of its first column \cite{toep2circApp2003}. For a Toeplitz matrix,
the most efficient algorithms for Cholesky factorization are Levinson
or Schur algorithms \cite{statDSP}, which involve $\mathcal{O}(N_{f}^{2})$
computations. In contrast, the eigen-decomposition of a circulant
matrix can be done efficiently using the fast Fourier transform (FFT)
and its inverse with only $\mathcal{O}\left(N_{f}\,log(N_{f})\right)$
operations. 

The preceding discussion shows that the problem of designing sparse
FIR equalizers can be cast into one of sparse approximation of a vector
by a fixed dictionary. The general form of this problem is given by
(\ref{eq:propFW}). To solve this problem, we use the well-known Orthogonal
Matching Pursuit (OMP) greedy algorithm \cite{omp07} that estimates
$\widehat{\boldsymbol{w}}_{s}$ by iteratively selecting a set $S$
of the sparsifying dictionary columns (i.e., atoms $\boldsymbol{\phi}_{i}'s$)
of $\boldsymbol{\varPhi}$ that are most correlated with the data
vector $\boldsymbol{b}$ and then solving a restricted least-squares
problem using the selected atoms. The OMP stopping criterion ($\rho$)
is changed here from an upper-bound on the residual error to an upper-bound
on the Projected Residual Error (PRE), i.e., ``$\boldsymbol{A}\times\mbox{Residual Error}$''.
The computations involved in the OMP algorithm are well documented
in the sparse approximation literature (e.g., \cite{omp07}) and are
omitted here due to page limitations. 

Unlike conventional compressive sensing \cite{CS}, where the measurement
matrix is a fat matrix, the sparsifying dictionary in our framework
is a square one with full rank. However, OMP and similar methods can
still be used if $\boldsymbol{R}_{yy}$ can be decomposed into $\boldsymbol{Q}\boldsymbol{Q}^{H}$
and the data vector $\boldsymbol{b}$ is compressible \cite{sparsefeng2012,sparseFilterDesign13}.
Among the proposed dictionaries shown in Table \ref{tab:Examples-of-different},
only $\boldsymbol{U}^{H}$ is not a valid choice of $\boldsymbol{\varPhi}$
since the data vector $\boldsymbol{b}$ associated with it can not
be compressed into a lower dimensional space without significant information
loss and, in addition, its PRE is large. Notice that it is better
to keep the PRE as small as possible to limit the amount of noise
in the data. 

Our next challenge is to determine the best sparsifying dictionary
for use in our framework. We know from the sparse approximation literature
that the sparsity of the OMP solution tends to be inversely proportional
to the worst-case coherence $\mu\left(\boldsymbol{\varPhi}\right)$,
{\small{}$\mu\left(\boldsymbol{\varPhi}\right)\triangleq\underset{i\neq j}{\mbox{max}}\frac{\left|\left\langle \phi_{i},\,\phi_{j}\right\rangle \right|\,}{\left\Vert \phi_{i}\right\Vert _{2}\left\Vert \phi_{j}\right\Vert _{2}}$}
\cite{finiteSparseFilter013,greedIsGood03}. Notice that $\mu\left(\boldsymbol{\varPhi}\right)\in\left[0,1\right]$.
Next, we investigate the coherence of the dictionaries proposed in
Table \ref{tab:Examples-of-different}.

\vspace{-1.5em}

\subsection{Worst-Case Coherence Analysis\label{sub:Preliminary-Analysis}}

We carry out a coherence metric analysis to gain some insight into
the performance of the proposed sparsifying dictionaries and the behavior
of the resulting sparse equalizers. First and foremost, we are concerned
with analyzing $\mu\left(\boldsymbol{\varPhi}\right)$ to ensure that
it does not approach $1$ for the proposed sparsifying dictionaries.
In addition, we are interested in identifying which $\boldsymbol{\varPhi}$
has the smallest coherence and, hence, gives the sparsest FIR equalizer.
We proceed as follows. We estimate an upper bound on the worst-case
coherence of $\boldsymbol{R}_{yy}$ and evaluate its closseness to
$1$. Then, through simulation we show that the coherence of other
dictionaries, which can be considered as the square roots of $\boldsymbol{R}_{yy}$
in the spectral-norm sense, i.e., $\left\Vert \boldsymbol{R}_{yy}\right\Vert _{2}=\left\Vert \boldsymbol{L}\boldsymbol{L}^{H}\right\Vert _{2}\leq\left\Vert \boldsymbol{\boldsymbol{L}\vphantom{\boldsymbol{L}^{H}}}\right\Vert _{2}^{2}$,
$\!\left\Vert \boldsymbol{R}_{yy}\right\Vert _{2}\!\leq\!\left\Vert \boldsymbol{\Lambda}^{\frac{1}{2}}\boldsymbol{\boldsymbol{P}}^{H}\right\Vert _{2}^{2}$
and $\left\Vert \boldsymbol{R}_{yy}\right\Vert _{2}\leq\left\Vert \boldsymbol{\boldsymbol{D}^{\frac{1}{2}}\boldsymbol{U}^{H}}\right\Vert _{2}^{2}$,
will be less than that of $\mu(\boldsymbol{R}_{yy})$. Interestingly,
$\boldsymbol{R}_{yy}$ has a well-structured (Hermitian Toeplitz)
closed form in terms of the CIR coefficients, filter time span $N_{f}$
and SNR, i.e., $\boldsymbol{R}_{yy}=\boldsymbol{H}\boldsymbol{H}^{H}+\mbox{\ensuremath{\frac{\mbox{1}}{SNR}}}\boldsymbol{I}$.
It can be expressed in a matrix form as 

\vspace{-2.5em}

{\small{}
\begin{equation}
\boldsymbol{R}_{yy}=\mbox{Toeplitz}\overbrace{\left(\left[\begin{array}{ccccccc}
r_{0} & r_{1} & \ldots & r_{v} & 0 & \ldots & 0\end{array}\right]\right)}^{\boldsymbol{\phi}_{1}^{H}}\,,\label{eq:R_yy_matrix_from}
\end{equation}
}where {\small{}$r_{0}={\displaystyle \sum_{i=0}^{v}\left|h_{i}\right|^{2}+\left(\mbox{SNR}\right)^{-1}}$,
$r_{j}=\sum_{i=j}^{v}h_{i}h_{i-j}^{*},\,\forall j\neq0$}. Assuming
high SNR, we can compute $\mu(\boldsymbol{R}_{yy})$ in terms of the
channel taps only. By noting that the columns of $\boldsymbol{R}_{yy}$
are fully defined by the first column, we can get the maximum possible
absolute inner product $\mu(\boldsymbol{R}_{yy})$ by simultaneously
maximizing the entries of $\boldsymbol{\phi}_{1},$ which results
in maximizing all columns entries accordingly. While we can pose the
problem of computing $\mu(\boldsymbol{R}_{yy})$ in terms of maximizing
the sum of the inner product $\left\langle \boldsymbol{\phi}_{i},\,\boldsymbol{\phi}_{j}\right\rangle ,\,\forall i\neq j$,
it turns out that it is equivalent to maximizing $r_{1}$ due to the
special structure of $\boldsymbol{R}_{yy}$. Hence, an upper bound
on $\mu(\boldsymbol{R}_{yy})$ in the high SNR setting can be derived
by solving the following optimization problem 

\vspace{-2em}

\begin{equation}
\mbox{max}\,\,\,\sum_{i=1}^{v}\left|h_{i}h_{i-1}^{*}\right|\,\,\,\,\,\,\,\mbox{s.t}.\,\,\,\,\,\,\,\sum_{i=0}^{v}\left|h_{i}\right|^{2}=1.\label{eq:optWorstTaps}
\end{equation}

The solution of (\ref{eq:optWorstTaps}) gives the worst CIR vector
$\boldsymbol{h}$ which is then used to estimate an upper-bound on
$\mu(\boldsymbol{R}_{yy})$ for any given channel length $v$. This
solution has a symmetric structure that can be obtained by solving
a simpler equivalent problem formulated as below

\vspace{-1.0em}

\begin{equation}
\mbox{max}\,\,\,\left|\boldsymbol{h}^{H}\boldsymbol{R}\boldsymbol{h}\right|\,\,\,\,\,\,\,\mbox{s.t}.\,\,\,\,\,\,\,\boldsymbol{h}^{H}\boldsymbol{h}=1\,,\label{eq:quadEqProb}
\end{equation}
where $\boldsymbol{h}=\left[\begin{array}{cccc}
h_{0} & h_{1} & \ldots & h_{v}\end{array}\right]^{H}$ is the length-$(v+1)$ CIR vector and $\boldsymbol{R}$ is a matrix
that has ones along the super and sub-diagonals. The solution of (\ref{eq:quadEqProb})
is the eigenvector corresponding to the maximum (or minimum, since
$\mu(\boldsymbol{R}_{yy})$ is defined in terms of absolute value)
eigenvalue of $\boldsymbol{R}$. Interestingly, the eigenvalues $\lambda_{s}$
and eigenvectors $h_{j}^{(s)}$ of the matrix $\boldsymbol{R}$ have
the following simple closed forms \cite{eigValueVector_R}

\vspace{-1.5em}{\small{}
\begin{eqnarray}
\lambda_{s} & = & 2\,\mbox{cos}(\frac{\pi s}{v+2})\,\,\,\,,\,\,\,\,h_{j}^{(s)}=\sqrt{\frac{2}{v+2}}\mbox{sin\ensuremath{(\frac{j\pi s}{v+2})\,,\,}}\label{eq:worst-taps}
\end{eqnarray}
}where $s,j=1,\ldots,v+1.$ Finally, by numerically evaluating $h_{j}^{(s)}$
for the maximum $\lambda_{s}$ we find that the worst-case coherence
of $\boldsymbol{R}_{yy}$ (for any $v$) is sufficiently less than
1, which points to the likely success of OMP in providing the sparsest
solution $\widehat{\boldsymbol{w}}_{s}$ which is corresponding to
the dictionary that has the smallest $\mu(\boldsymbol{R}_{yy})$.
Next, we will report the results of our numerical experiments to evaluate
the performance of our proposed framework under different sparsifying
dictionaries. 

\vspace{-1.0em}

\section{Simulation Results \label{sec:Simulation-Results}}

\begin{figure}
\vspace{-2em}

\includegraphics[scale=0.28]{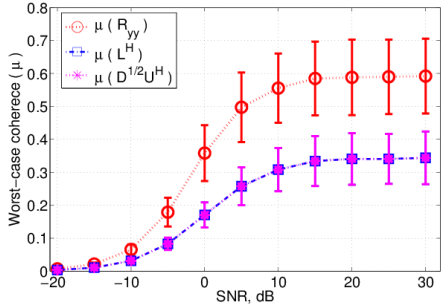}

\vspace{-1em}\protect\caption{{\footnotesize{}Worst-case coherence for the sparsifying dictionaries
versus input SNR. Each point represents the mean of 5000 channel realizations.}\label{fig:mu_versus_snr}}

\vspace{-1em}
\end{figure}

\begin{figure}
\vspace{-0.45em}

\includegraphics[scale=0.28]{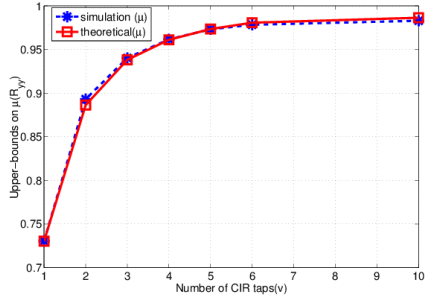}

\vspace{-1em}

\protect\caption{{\footnotesize{}Upper-bounds on $\boldsymbol{R}_{yy}$ worst-case
coherence versus channel length under unit-energy channel constraint.
}\label{fig:upperBound}}

\vspace{-2em}
\end{figure}

Throughout the simulations, the used CIRs are unit-energy symbol-spaced
FIR filters with $v$ nonzero taps generated as zero-mean uncorrelated
complex Gaussian random variables. We assume $v=5$ and $N_{f}=35$
\cite{jcioffi}. To quantify the performance of the sparsifying dictionaries
involved in our analysis in terms of coherence, we plot the worst-case
coherence versus the input SNR in Figure \ref{fig:mu_versus_snr}.
Note that a smaller value of $\mu\left(\boldsymbol{\varPhi}\right)$
indicates that a reliable sparse approximation is more likely. Clearly,
$\boldsymbol{R}_{yy}$ has higher $\mu\left(\boldsymbol{\varPhi}\right)$
which reflects higher similarities between its columns compared to
$\boldsymbol{D}^{\frac{1}{2}}\boldsymbol{U}^{H}$ and $\boldsymbol{L}^{H}$
(both have the same $\mu\left(\boldsymbol{\varPhi}\right)$). The
coherence increases with SNR up to a certain limit and then saturates.
This can be interpreted by the fact that, at high SNR, the noise effects
are negligible and, therefore, the sparsifying dictionaries (e.g.,
{\small{}$\boldsymbol{R}_{yy}\approx\boldsymbol{H}\boldsymbol{H}^{H}$})
do not depend on the SNR. Hence, the coherence converges to a constant.
In contrast, at low SNR, the noise effects dominate the channel effects.
Hence, the channel can be approximated as a memoryless (i.e., 1 tap)
channel. Then, the dictionaries (e.g., {\small{}$\boldsymbol{R}_{yy}\approx\frac{1}{SNR}\boldsymbol{I}$})
can be approximated as a multiple of the identity matrix, i.e., $\mu\left(\boldsymbol{\varPhi}\right)\rightarrow0$.
Figure \ref{fig:upperBound} shows theoretical bounds, estimated through
(\ref{eq:worst-taps}), and empirical upper bounds on the worst-case
coherence $\mu\left(\boldsymbol{R}_{yy}\right)$. This figure shows
that the maximum coherence is sufficiently less than 1 and the mismatch
between the theoretical and simulation results is negligible (only
0.67\%). 

We further compare the sparse FIR equalizer designs based on the dictionaries
$\boldsymbol{D}^{\frac{1}{2}}\boldsymbol{U}^{H}$, $\boldsymbol{L}^{H}$
and $\boldsymbol{R}_{yy}$, denoted as $\boldsymbol{w}_{s}(\boldsymbol{D}^{\frac{1}{2}}\boldsymbol{U}^{H})$,
$\boldsymbol{w}_{s}(\boldsymbol{L}^{H})$ and $\boldsymbol{w}_{s}(\boldsymbol{R}_{yy})$,
respectively, to study the effect of $\mu$ on their performance.
The case of $\boldsymbol{\varPhi}=\boldsymbol{\Lambda}^{\frac{1}{2}}\boldsymbol{\boldsymbol{P}}^{H}$
is not presented here since its performance is almost equivalent to
{\small{}$\boldsymbol{L}^{H}$}. The OMP algorithm is used to compute
the sparse approximations. The OMP stopping criterion is set to be
a function of the PRE such that: {\small{}Performance Loss ($\eta$)$=10\,\mbox{Log}_{10}\left(\frac{SNR(\boldsymbol{w}_{s})}{SNR(\boldsymbol{w}_{opt})}\right)\leq10\,\mbox{Log}_{10}\left(1+\frac{_{\delta_{eq}}}{\xi_{m}}\right)\triangleq\eta_{max}$}.
Here, $\delta_{eq}$ is computed based on an acceptable $\eta_{max}$
and, then, the coefficients of $\widehat{\boldsymbol{w}}_{s}$ are
computed through (\ref{eq:propFW}). The percentage of the active
taps is calculated as the ratio between the number of nonzero taps
to the total number of filter taps, i.e., $N_{f}$. For the MMSE equalizer,
where none of the coefficients is zero, the number of active filter
taps is equal to the filter span. The decision delay is set to be
{\small{}$\Delta\approx\frac{N_{f}+v}{2}$}\cite{jcioffi}. 
\begin{figure}[t]
\vspace{-2em}

\includegraphics[scale=0.29]{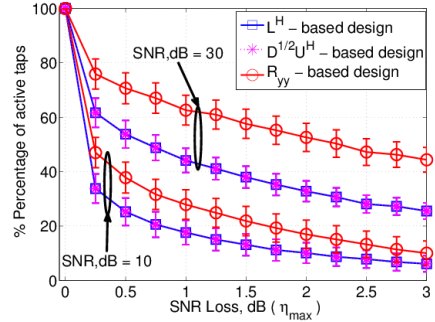}

\vspace{-1em}

\protect\caption{{\footnotesize{}Percentage of active taps versus the performance loss
($\eta_{max}$) for the sparse LEs (5000 channel realizations).}\label{fig:activeTaps_versus_eta_max}}

\vspace{-1.1em}
\end{figure}
\begin{figure}
\includegraphics[scale=0.29]{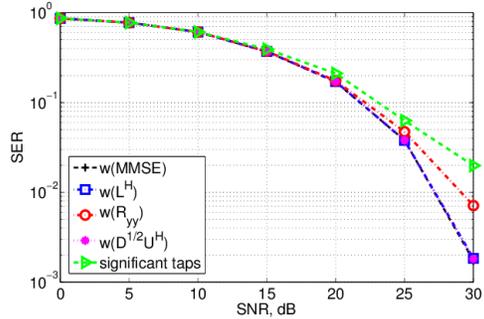}

\vspace{-1.2em}\protect\caption{{\footnotesize{}SER comparison between the MMSE non-sparse LE, the
proposed sparse LEs $\boldsymbol{w}_{s}(\boldsymbol{D}^{\frac{1}{2}}\boldsymbol{U}^{H})$,
$\boldsymbol{w}_{s}(\boldsymbol{L}^{H})$, $\boldsymbol{w}_{s}(\boldsymbol{R}_{yy})$
and the ``significant-taps'' based LE with sparsity level = 0.25
and 16-QAM modulation.}\label{fig:BER_versus_SNR}}

\vspace{-1.5em}
\end{figure}
 \vspace{-1em}

Figure \ref{fig:activeTaps_versus_eta_max} plots the percentage of
the active taps versus the performance loss $\eta_{max}$ for the
proposed sparse FIR-LEs. We observe that a lower active taps percentage
is obtained when the coherence of the sparsifying dictionary is small.
For instance, allowing for $0.25$ dB SNR loss results in a significant
reduction in the number of active LE taps. Approximately two-thirds
(two-fifths) of the taps are eliminated when using $\boldsymbol{w}_{s}(\boldsymbol{D}^{\frac{1}{2}}\boldsymbol{U}^{H})$
and $\boldsymbol{w}_{s}(\boldsymbol{L}^{H})$ at SNR equals to 10$\,$(30).
The sparse LE $\boldsymbol{w}_{s}(\boldsymbol{R}_{yy})$ needs more
active taps to maintain the same SNR loss as that of the other sparse
LEs due to its higher coherence. This suggests the smaller the worst-case
coherence, the sparser is the equalizer. Moreover, a lower sparsity
level (active taps percentage) is achieved at higher SNR levels which
is consistent with the previous findings (e.g., in \cite{tapPositions07}).
Furthermore, reducing the number of active taps decreases the filter
equalization design complexity and, consequently, the power consumption
since a smaller number of complex multiply-and-add operations are
required. 

In Figure \ref{fig:BER_versus_SNR}, we compare the symbol error rate
(SER) performance of our proposed sparse LEs with the proposed approach
in \cite{strongestTap} which we refer to it as the ``significant-taps''
approach. In that approach, all of the MMSE LE taps are computed and
only the $K$ significant ones are retained. Assuming a $25\%$ sparsity
level, both the $\boldsymbol{w}_{s}(\boldsymbol{D}^{\frac{1}{2}}\boldsymbol{U}^{H})$
and $\boldsymbol{w}_{s}(\boldsymbol{L}^{H})$ sparse LEs achieve the
lowest SER followed by $\boldsymbol{w}_{s}(\boldsymbol{R}_{yy})$,
while the ``significant-taps'' performs the worst. In addition to
this performance gain, the complexity of the proposed sparse LEs is
less than that of the ``significant-taps'' LE since only an inversion
of an $N_{s}\times N_{s}$ matrix is required (not $N_{f}\times N_{f}$
as in the ``significant-taps'' approach) where $N_{s}$ is the number
of nonzero taps. Although the $\boldsymbol{w}_{s}(\boldsymbol{D}^{\frac{1}{2}}\boldsymbol{U}^{H})$
and $\boldsymbol{w}_{s}(\boldsymbol{L}^{H})$ LEs achieve almost the
same SER, the former has a lower decomposition complexity since its
computation can be done efficiently with only the FFT and its inverse. 

\vspace{-1.5em}

\section{Conclusions\label{sec:Conclusion-and-Future}}

\vspace{-0.5em}

In this paper, we proposed a general framework for sparse FIR equalizer
design based on a sparse approximation formulation using different
dictionaries. In addition, we investigated the coherence of the proposed
dictionaries and showed that the dictionary with the smallest coherence
gives the sparsest equalizer design. The significance of our approach
was shown analytically and quantified through simulations. 

\balance

\vspace{-1.0em}

\bibliographystyle{IEEEtran}
\bibliography{globaSIPRef}

\end{document}